# A PROBLEM WITH THE CLUSTERING OF RECENT MEASURES OF THE DISTANCE TO THE LARGE MAGELLANIC CLOUD

Bradley E. Schaefer

Department of Physics & Astronomy
Louisiana State University
Baton Rouge, LA 70803
schaefer@lsu.edu

**ABSTRACT**

The distance to the Large Magellanic Cloud (LMC) has long been of key importance for the distance ladder and the distances to all galaxies, and as such many groups have provided measurements of its distance modulus (µ) with many methods and various means of calibrating each method. Before the year 2001, the many measures spanned a wide range (roughly $18.1 < \mu < 18.8$) with the quoted error bars being substantially smaller than the spread, and hence the consensus conclusion being that many of the measures had their uncertainties being dominated by unrecognized systematic problems. In 2001, the Hubble Space Telescope Key Project (HSTKP) on the distance scale made an extensive analysis of earlier results and adopted the reasonable conclusion that the distance modulus is 18.50±0.10 mag, and the community has generally accepted this widely popularized value. After 2002, 31 independent papers have reported new distance measures to the LMC, and these cluster tightly around µ=18.50 mag. Indeed, these measures cluster *too* tightly around the HSTKP value, with 68% of the measures being within 0.5-sigma of 18.50 mag. A Kolmogorov-Smirnov test proves that this concentration deviates from the expected Gaussian distribution at a >3-sigma probability level. This concentration is a symptom of a worrisome problem. Interpretations considered include correlations between papers, widespread over-estimation of error bars, and band-wagon effects. This note is to alert workers in the field that this is a serious problem that should be addressed.

# 1. BACKGROUND

The distance modulus ($\mu$) of the Large Magellanic Cloud (LMC) has long been a key question for the distance ladder and all extragalactic distances. For the Hubble Space Telescope Key Project (HSTKP) to measure the Hubble constant ($H_0$), "the uncertainty in the distance to the LMC is one of the largest remaining uncertainties in the overall error budget for the determination of $H_0$" (Freedman et al. 2001). If the LMC $\mu$ is changed by 0.1 mag, then the $H_0$ value will change by 5%.

Hundreds of papers have been published which present independent measurements of the LMC distance. Recent compilations of these values appear in Westerlund (1997), Cole (1998), Gibson (2000), Freedman et al. (2001), Benedict et al. (2002a), Clementini et al. (2003), Walker (2003), and Alves (2004). These have used many different standard candles (including Cepheids, RR Lyrae stars, SN1987A, eclipsing binaries, the tip of the red giant branch, the red clump, and Mira variables), which have each been calibrated with many independent methods. For example, Cepheids have been calibrated with the Baade-Wesselink method, main sequence fitting in galactic clusters, nonlinear pulsation modeling, *Hipparcos* parallaxes, and *HST* parallaxes.

Before the year 2001, the published $\mu$ values were spread from 18.1 to 18.8 mag with the stated error bars being much smaller than the spread. This was generally regarded as a symptom that many measures had large unrecognized systematic uncertainties. For example, "It is clear from the wide range of moduli compared to the quoted internal errors in Figure 5 [of the quoted paper] that systematic errors affecting individual methods are still dominating the determinations of LMC distances." (Freedman et al. 2001). Gibson (2000) concluded "it is clear from even a cursory examination of [his] Figure 1 that significant unappreciated systematic uncertainties exist in many published values of $\mu_{LMC}$".

In 2001, the final report of the HSTKP (Freedman et al. 2001) made a detailed analysis of the earlier LMC distance measures and concluded (based largely on Madore & Freedman 1991) that $\mu=18.50\pm0.10$. This conclusion is quite reasonable and has been widely publicized and adopted as a consensus of our community.

After 2001, the literature still contains many reported new measures of the LMC distance by many different methods. But suddenly, the wide scatter has disappeared. Alves (2004) is the only compilation that reports on post-HSTKP measures alone, and he finds that "The average of 14 recent measurements of [$\mu$] implies a true distance modulus of 18.50±0.02 mag, and demonstrates a trend in the past two years of convergence towards a standard value." He even goes so far as to say "*Regarding the convergence of published LMC distance results, I suggest to you that the fat lady has begun to sing*." [his italics]. Popular press reports repeat the claim that there is "a trend in the last two years of, finally, convergence toward this standard value." (MacRobert 2004).

But no one has offered any explanation as to why the situation should *suddenly* improve from one where most methods were being dominated by large systematic errors to one where all these same methods have their systematic errors essentially vanish. Equally disturbing is the fact that Alves' compilation shows 14 values that are *too* consistent with the HSTKP value. That is, *all* 14 values have 18.50 within their quoted 1-sigma error bars. A chi-square test for the hypothesis that $\mu=18.50$ has a reduced $\chi^2$ of 0.23 (with 13 degrees of freedom), which is improbably low (with a probability of

0.0022) if the error estimates are correct. (Alves only added one sentence saying "The reduced chi-square is less than one, which suggests that the adopted error bars may be too conservative" after I had pointed out the problem as it appeared in his preprint.) I next performed a cursory literature search and quickly turned up seven more measures, with the entire list of 21 measures having a reduced $\chi^2$=0.19 and a chance probability of such a low value of 0.00003. This is a disturbing situation, so I have made a thorough analysis, and this is what I report on in this paper.

## 2. LMC DISTANCE MEASURES

I have collected published measures of the LMC distance modulus that have appeared after 1 January 1997. For dates before 2002, most of the publications were previously compiled by Benedict et al. (2002) and Clementini et al. (2003), with these not being listed separately in this paper. My search for additional measures was made with the ADS database (http://adsabs.Harvard.edu/), the astro-ph preprint server (http://www.arxiv.org/astro-ph/), and the indexes of various journals. I suspect that I have not found all reported measures of $\mu$, but likely there are only a few missing, and the completeness is not correlated with the reported $\mu$ value. In all, I have found 45 papers that report on $\mu$ with error bars, and these are tabulated in Table 1. An additional 99 values were collected from the Benedict et al. (2002) and Clementini et al. (2003) papers. The first column gives a date for each paper, usually its acceptance date. The second and third columns cite the reference and a short identification of the method used. The fourth column gives the quoted $\mu$ value and its 1-sigma uncertainty ($\sigma$). The fifth column gives the deviation of the reported $\mu$ value from that of the HSTKP value (18.50 mag) divided by the 1-sigma uncertainty, that is $D=(\mu-18.50)/\sigma$. If the true $\mu$ value is 18.50 mag and if the various reported uncertainties are correct, then D should be a Gaussian distribution centered at zero with an RMS scatter of unity.

A variety of minor dilemmas occur when trying to summarize a long paper into one number for Table 1. For example, in Fitzpatrick et al. (2003), the distance to the eclipsing binary HV5936 is given as $\mu$=18.18±0.09, but they then correct this distance to the middle of the LMC galaxy as reported in Table 1. Benedict et al. (2002a) reports their value as 18.38-18.53 with error bars of -0.11 and +0.10, while I have represented this as 18.455±0.13. Keller & Wood (2002; 2006) only include one source of uncertainty into their very small quoted error bars (±0.02 mag and ±0.018 mag respectively) despite discussing additional uncertainties that would greatly increase the error bars, yet I have used their published error bar. Bono et al. (2002b) presents two results (18.53±0.08 based on a theoretical calibration and 18.48±0.13 based on an empirical calibration) that I have combined as 18.51±0.11 in Table 1. Sebo et al. (2002) report on values for two cuts for which their best result (de-reddened and with a plausible galaxy tilt) are 18.53 and 18.54 mag, but no error bars are quoted so this measure was not included in my Table 1.

Some of the papers report multiple values for $\mu$ based on alternative reasonable assumptions or ranges of acceptable inputs. (As pointed out in the last section of this paper, this is a good practice.) Should these multiple values be presented in Table 1 as individual lines or as some combined measure? With the alternative $\mu$ values usually being based on analyses that differ by only a small fraction of their input, the values are highly non-independent, and as such they should not be separately listed in Table 1. Rather, the range of alternative values is an expression of the systematic errors not included in the quoted error bars. As such, this systematic uncertainty should be added in quadrature with the quoted error bars to yield a total uncertainty for inclusion in Table 1.

## 3.  AFTER 2002

Alves (2004) claims that the post-HSTKP measures of the LMC distance have converged. But is this convergence *too* good?

A quick scan through Table 1 shows many cases where the measured value is close to the HSTKP value. Of the 31 measures dated after 2002, all but one (i.e., 97%) are within 0.10 mag of the HSTKP value. (This is to be compared to the 46% between 18.40 and 18.60 for the pre-HSTKP compilation of 84 papers from 1997-2000 [Benedict et al. 2002a]. This demonstrates the dramatic change around the time of the publication of the HSTKP results.) A formal weighted average of the post-2002 measures returns a value of 18.51±0.01 mag (an unweighted average gives 18.49±0.01 mag), which is very close to the HSTKP value.

The disturbing property of Table 1 is the over-abundance of measures that agree with the HSTKP to much greater accuracy than the quoted error bars. For the post-2002 measures, 87% (27 out of 31) of the measures have the reported $\mu$ value within 1-sigma of the HSTKP value (i.e., $|D|\leq1$), whereas the expected fraction is 68%. (Indeed, the fraction would be 90% had not Keller & Wood quoted an artificially low error bar.) Also, 68% (21 out of 31) of the measures have the reported $\mu$ value within 0.5-sigma of the HSTKP value (i.e., $|D|\leq0.5$), whereas the expected fraction is 38.3% (12 out of 31). These numbers are startling deviations from the expectations.

A chi-square analysis can be used to test whether the overall spread of the deviations (D) is as expected for a model that the true $\mu$ value is that given by the HSTKP (18.50 mag). The $\chi^2$ value will simply be the sum of the squares of the D values in the last column of Table 1. The number of degrees of freedom equals the number of reported measures (31) as there are no adjustable model parameters since I calculated D with the model of the HSTKP $\mu$ rather than the weighted average of the values. I find that $\chi^2$=20.3 for a reduced chi-square of 0.65. This value is not exceedingly small, being surprising at only the two-sigma level. (If the Keller & Wood (2006) measure is excluded due to their not including their stated large systematic errors, then the reduced chi-square becomes 0.51, with this being improbable at the 1% probability level.)

However, the chi-square test is not the most sensitive test for the question of whether the reported values cluster too tightly around the HSTKP value. The reason is that we already know that historically there are large systematic errors unaccounted for (cf. Benedict et al. 2002a) and just a few of these included in the post-2002 sample will provide a high chi-square contribution that would mask an anomalous concentration. An additional reason is that any exclusion of systematic uncertainties will also provide an erroneous high chi-square contribution. Indeed, just three of the post-2002 measures (one with an admitted low error bar of ±0.018 mag) provides 14.0 out of the total $\chi^2$=20.3.

A more sensitive test than the chi-square for the question at hand is the Kolmogorov-Smirnov (K-S) test (Press et al. 1986). Here, the observed distribution of differences from the HSTKP value is compared to the ideal distribution arising if the LMC distance measures have a Gaussian distribution with the quoted error bars. As the deviations from the Gaussian shape are symmetric, I will examine the observed distribution of |D| as this combines any effects of clustering on both the positive and

negative sides. This K-S test is quite general in that it makes no assumption as to the fraction or size-scale in the cluster, and the effects of outliers is minimal.

The K-S test starts with the construction of cumulative distributions where the fraction of values less than |D| are tabulated from small to large |D|. For small |D| the fraction will be near zero while for large |D| the fraction will approach unity. Two such cumulative distributions are constructed, one for the 31 observed post-2002 values and the other for the Gaussian model (see Figure 1). With normal fluctuations, the two cumulative distributions will deviate from each other, and they will have some maximum difference. The greater the maximum difference is, the lower the probability that the observed distribution is correctly modeled by the theoretical distribution. The maximum difference between the two cumulative distributions of |D| is 0.33 at |D|=0.59. This large of a difference is unlikely to occur, with the probability of good data and a good model producing such a large difference equal to 0.0023. (With the exclusion of the Keller & Wood report [for the cause of its artificially small error bar], the probability is reduced to 0.0010.) The probability of 0.0023 corresponds to worse than a 3.0-sigma discrepancy, and constitutes strong evidence that the observed distribution does not match the model.

So here we have it, the post-2002 measures of $\mu$ are *too* tightly clustered to the HSTKP value (compared to what is possible given the quoted error bars) with a >3-sigma confidence level. There are only two ways that $D=(\mu-18.50)/\sigma$ can be systematically near zero, and that is either if the published $\mu$ values were artificially selected/adjusted to be near 18.50 mag or if the published $\sigma$ values are systematically too *large* by about a factor of two. Either possibility strikes at the heart of the reliability of post-2002 reports.

## 4. 1997-PRESENT

For understanding the artificial clustering of LMC distances, it is helpful to track how this clustering has evolved in time. We already know that the measures before ~2001 have large unrecognized systematic errors while the measures after ~2002 have a clustering too tight to be consistent with their reported errors. But exactly when did this transition occur and how long did it take?

To answer these questions, I have assembled all the published measures of LMC distances from 1997 to present. These come from Benedict et al. (2002), Clementini et al. (2003) and Table 1. These measures of $\mu$ are plotted as a function of date in Figure 2. In addition, the values of |D| are plotted as a function of date in Figure 3.

These 144 measures have also been divided into various time intervals, with their statistical properties tabulated in Table 2. The first column gives the years or year ranges for the input. The second column gives the number of published values in the date range. The third and fourth columns give the unweighted average and the RMS scatter for the reported $\mu$ values in the time interval. The last column gives the median |D| value.

From 1997 to 2001, the average LMC distance modulus, $<\mu>$, is bouncing up and down from 18.33 to 18.55. After 2002, $<\mu>$ is relatively constant (18.46-18.52) and close to the HSTKP value. Apparently, the transition occurred around the end of 2001.

From 1997 to 1999, the RMS scatter in the reported $\mu$ values was relatively large (i.e., ≥0.13 mag). From 2003 to present, the RMS scatter is relatively low (i.e., ≤0.06 mag). Based on this measure, the transition occurred between 2000 and 2002, with the RMS scatter having intermediate size (i.e., 0.10 to 0.11 mag).

The median |D| value is a reasonable measure of the degree of too-tight clustering towards the central value. This measure is relatively insensitive to the existence of outliers and it can be directly compared to the expectation for a good Gaussian distribution. (for which the median |D| should be 0.68). From 1997 to 2002, the median |D| is larger than the value expected for good Gaussian errors. From 2003 to present, the median |D| is greatly smaller than expected for good Gaussian errors. From this, it looks like the transition occurred in 2002.

From these three statistics, it appears that the transition from widely-scattered to tightly-clustered (when compared to the quoted error bars) occurred sometime around 2002. The data do not have enough resolution in time to set tight limits on the duration of the transition, so the transition might be fast and sometime in 2002 or it might be perhaps a two years long interval from around 2001 to 2002.

The timing and duration of the transition allows us to reject some possible causes. For example, various of the substantial advances in data that have widespread application amongst all the methods do not correspond in time to the observed transition. In particular, the sudden availability of *Hipparcos* parallaxes and the OGLE and MACHO photometry both happened long before the year 2000 and hence cannot be responsible for the transition. In addition, the relatively short duration of the transition indicates that the transition is not caused by the steady accumulation of improvements in methods, whether applicable to individual techniques or applicable to multiple techniques. The short unresolved transition is consistent with some single cause dating to around 2001-2002.

The HSTKP paper was submitted on 2000.6, accepted on 2001.0, and published on 2001.4. The tremendous publicity for the HSTKP was throughout 2001. If the transition was caused by the HSTKP, then the transition should appear sometime around 2001 plus the average time for new research to be performed plus the average time resulting from the publication process. As such, we have a prediction that if the transition is caused by the HSTKP, then the transition should last for a year or so starting sometime in 2001.

We have a close match between the observed transition date (2002 or perhaps 2001 to 2002) and the predicted transition date (2001 to 2002) if the HSTKP is a primary cause of the transition. This coincidence in time does not prove a causal connection. However, given that we know independently about the massive publicity and the instant 'consensus' caused by the HSTKP, the coincidence in time does provide reasonable evidence that the transition is related to the HSTKP.

## 5. THE CEPHEID DISTANCES

The 31 post-2002 measures of the LMC distance use six classes of standard candles, with many different methods for calibrating each. Nevertheless, there are certainly overlap in data, methods, and assumptions between the papers. Are the 31 measures largely independent? If not, then the effective number of degrees of freedom used in Section 3 might be substantially smaller than 31. Another way to look at this is that the existence of substantial correlations amongst the 31 measures could cause individual reports to be clustered more tightly than their quoted error bars.

The most commonly used class of standard candles in the 31 post-2002 papers is the Cepheids, with 13 papers. This is the case where correlations would be the most pervasive. So in this Section, I will examine in detail the correlations amongst the measured LMC distances based on Cepheids.

I have examined the 13 post-2002 Cepheid papers and I find that they have little overlap. The papers calibrate the absolute magnitude of Cepheids by many different methods, including theoretical models, parallaxes, the Baade-Wesselink method, the hydrogen maser distance to NGC4258, and main sequence fitting of galactic clusters with Cepheids. This very wide range of calibration methods provides confidence that any correlations will be substantially restricted. To provide a detailed analysis, I will closely describe the relations between the groups of Cepheid papers that share the same calibration methods.

The first group is the three papers that calibrate their LMC Cepheid distances with theoretical models (Moskalik & Dziebowski 2005; Keller & Wood 2006; Testa et al. 2007). In all cases, the theoretical models are completely independent by widely separated groups. Indeed, the models are for completely different classes of Cepheids, with the three papers modeling triple-mode Cepheids, bump Cepheids, and normal Cepheids respectively. The input light curves of the LMC Cepheids were taken from OGLE-II I-band data, MACHO BR data, and Las Campanas JK data respectively. There is no overlap between the target stars. As such, these three apparently similar papers have no correlation.

The second group is the two papers (Abrahamyan 2004; Benedict et al. 2007) that calibrate the Cepheid absolute magnitudes with parallaxes of galactic Cepheids. The basic inputs are completely different as Abrahamyan uses *Hipparcos* parallaxes while Benedict et al. use *HST* parallaxes. On top of that, the two papers use completely different LMC data and substantially different analysis assumptions. So despite the use of a few shared targets for calibration, the data and analyses are completely independent.

The third group is the three papers (Storm et al. 2004; 2006; Geiren et al. 2005) that all share a core of collaborators and the same technique (the Baade-Wesselink method) to calibrate the LMC Cepheid distances. These three papers have greatly different input. Storm et al. (2004) apply the Baade-Wesselink method to galactic Cepheids so as to calibrate a P-L relation, Gieren et al. (2005) apply the Baade-Wesselink method to 13 Cepheids across the LMC bar, and Storm et al. (2006) apply the Baade-Wesselink method to 6 Cepheids in the LMC cluster NGC1866. The same three papers use completely independent input data in the VIWJHK, VIWJK, and K bands, respectively. As for assumptions, these three papers use a metallicity correction but no projection correction, no metallity correction but with a projection correction (p=1.58-0.15 logP), and no metallicity correction but with a different projection correction (p=1.39-0.03 loP), respectively. In all, these three papers are completely independent in input.

Nevertheless, the 13 post-2002 Cepheid papers do carry some specific overlap. The paper of Storm et al. (2004) simply reports the result of Fouque, Storm, & Gieren (2003) plus a correction for metallicity effects, and so their two results are clearly strongly correlated. Both Fouque, Storm, & Gieren (2003) as well as Macri et al. (2006) adopt the OGLE period-luminosity relations as their LMC information, although they have greatly different methods of calibration. Ngeow & Kanbur (2007) base their calibration on the Geiren et al. (2005) distances from the Baade-Wesselink method, but they also use 630 other LMC Cepheid V-band light curves from OGLE and completely different methods and assumptions for analysis. These three cases are the only significant overlap that I can find. So, out of the 13 Cepheid papers, only two (Fouque, Storm, & Geiren 2003; Ngoew & Kanbur 2007) are not completely independent from each other.

A detailed examination of all the other post-2002 papers shows no significant overlap of data or targets. Out of all the 31 papers, I can only point to two that have any substantial overlap with other papers. In all, I conclude from these detailed analyses of the post-2002 papers that correlations are indeed present but only at a rather small level.

An alternative test for paper-to-paper correlations is to look at the median |D| values for just the Cepheid papers from 1997 to present. Any correlations amongst the Cepheid papers will certainly be stronger than the correlations amongst all 31 papers. As such, if the correlations are significant, then the Cepheid median |D| values should be more tightly clustered than those for all papers. If the correlations are not significant, then the Cepheid median |D| values should be similar to those of all the papers. So we have a test for the significance of correlations in creating the too-tight clustering reported in Section 3.

I have separated out the Cepheid-based measures from the entire data base. This results in 10, 13, and 13 measures in the year ranges of 1997-1999, 2000-2002, 2003-2007. The median |D| values are 1.03, 0.60, and 0.55 for before/during/after the transition. This is to be compared to 1.24, 0.76, and 0.43 respectively for all papers (see Table 2). The typical uncertainty in the median |D| for the Cepheids will be roughly ±0.3 (the RMS scatter in |D| divided by the square root of the number of items). As such, the two series of median |D| values are consistent. This is interpreted as implying that paper-to-paper correlations are small and do not cause the basic dilemma from Section 3.

## 6. SMALL MAGELLANIC CLOUD DISTANCES

Further insight into the cause of the too-tight clustering of LMC distances can be obtained by comparison with the situation for the Small Magellanic Cloud (SMC) distance modulus ($\mu_{SMC}$). The task of measuring the distance to the SMC is largely identical with that of measuring the LMC distance. As such, if there are any systematic advances that cause the clustering of LMC distances then these same advances should also apply to the SMC distances resulting in a similar too-tight clustering. Alternatively, if the LMC distance clustering is caused by 'sociological' influences connected with the HSTKP, then the SMC distances should display little clustering. As such, we have a means of distinguishing the cause of the transition.

Unfortunately, the literature contains much fewer reports of SMC distances than of LMC distances. (The reasons are the SMC is not the lynchpin for the entire extragalactic distance scale, its slightly farther distance makes it somewhat harder to get adequate observations, and its smaller mass means that there are fewer targets.) From 1991 to present, I find only 29 papers that report a derived SMC distance with error bars. I have divided them into time intervals 1991-1999 (with 14 measures), 2000-2002 (with 10 measures), and 2003-2007 (with 5 measures) so as to correspond to before/during/after the LMC transition.

The average (and median and weighted average) is close to $\mu_{SMC}$=18.9 mag. So I take the D value for the SMC to be $D_{SMC}=(\mu_{SMC}-18.9)/\sigma$. The RMS scatter of $\mu_{SMC}$ is

0.21, 0.14, and 0.11 mag for before/during/after the LMC transition. The median value of $|D_{SMC}|$ is 1.0, 1.0, and 1.2 for before/during/after the LMC transition.

If the LMC transition (with the median |D| going from 1.24 to 0.43) is caused by advances in techniques, then we would expect the median $|D_{SMC}|$ to undergo a similar transition. But this is *not* seen, as we have the median $|D_{SMC}|$ always ≥1 and even increasing somewhat after the transition. We do see a long-term improvement in the RMS scatter of the SMC distances (perhaps caused by the various real advances and improvements over the last decade), but this is *not* a clustering with respect to the reported error bars.

If the LMC transition is caused by 'sociological' effects connected with the widespread publicity for the HSTKP, then we would expect little effect on the SMC distances. The reason is that the SMC was not included in the HSTKP. Indeed, as predicted by this possibility, the $|D_{SMC}|$ does not undergo a transition from ≥1 to ≤0.5. If anything, $|D_{SMC}|$ grows even larger after the LMC transition. As such, the SMC distance measures provide reasonable evidence that the cause of the LMC transition is somehow associated with the HSTKP.

## 7. POSSIBLE INTERPRETATIONS

We are faced with a dilemma because the post-2002 distance measures for the LMC are greatly more concentrated around the HSTKP value than is plausible given the published error bars. Here are three possible interpretations:

### 7.1. CORRELATIONS BETWEEN PAPERS

Perhaps the post-2002 values are strongly correlated because they use the same sources for calibrations or share the same data. That is, if individual published measures largely use the same sources/data/methods/assumptions, then they should cluster more strongly than their reported error bars. For 31 post-2002 measures, the 13 that use Cepheids might share substantial amounts of data and methods, and similarly the 12 that use RR Lyrae stars also might share substantial amounts of data and methods. With strongly correlated analysis and input, the effective number of independent measures might be substantially less than 31, and then the concentration towards small values for |D| becomes less significant.

This interpretation has a variety of problems that make it hard for this explanation to explain the concentration to low |D|: (A) The exact same correlations are present in the pre-HSTKP measures, yet these measures are widely scattered by over 0.7 mag. No substantial advances in the many methods have occurred during the transition in 2001-2002, so the sudden drastic concentration to the HSTKP value cannot be due to correlations. (B) A detailed analysis of the 13 papers that use Cepheids shows that they actually use almost entirely separate data and methods for their calibration of the Cepheid period-luminosity relation as well as entirely different data and analysis for their measures of LMC Cepheids. (C) A detailed analysis shows that 29 out of the 31 post-2002 papers are completely independent. (D) The time behavior of the median |D| for the Cepheid papers is the same as that for all papers, and this points to any paper-to-paper correlation being insignificant. (E) The median $|D_{SMC}|$ does not share the sudden concentration to small values, and this is in contradiction to the idea that LMC distances are significantly correlated because the SMC distances would suffer the same correlations.

In all, the correlations between the papers is greatly too small to account for the too-strong clustering around the HSTKP value. We have many good and quantitative reasons to reject this first explanation for the too-tight clustering of LMC distances.

### 7.2. MOST QUOTED ERROR BARS HAVE LARGE OVER-ESTIMATES

Perhaps the majority of the LMC distances have published uncertainties that are substantially *over*-estimated? If so, then the |D| values in Table 1 will be larger than they should be and the discrepancy noted in Section 3 will be eased. With 68% of the measures having |D|≤0.5 (whereas 68% should have |D|≤1), a possible solution is to postulate that the published error bars are over-estimated on average by a factor of two.

This interpretation has a variety of problems: (A) The pre-HSTKP measures have published error bars that can only be substantially smaller than is realistic as evidenced by their large scatter about any single value when compared to their stated uncertainty, yet nothing has changed for all the methods around the 2001-2002 transition, so there is good reason to think that the post-HSTKP values will largely share the same under-estimation of uncertainties. (B) The reason for the common *under*-estimation of the true error bars is the presence of systematic errors whose effects are not included in the published error bars. This is what happened before the HSTKP. But to *over*-estimate error bars after the HSTKP, the majority of researchers would at least have to somehow suddenly eliminate or account for all systematic problems. This possibility is rather unlikely, and indeed, many of the post-2002 papers point explicitly to unresolved systematic errors that were not included into their published error bar. (C) The majority of individual investigators would *also* have to *over*-estimate their statistical errors by a factor of two on average. This is unlikely in even a small number of cases because methods of measuring statistical errors are well known and robust, while investigators take pride in designing a study that minimizes the size of their error bars that they will not then gratuitously over-estimate.

In all, there is no realistic hope that this second explanation (that the average experimenter suddenly started reporting error bars too big by a factor of two on average) can account for the too-tight clustering in the post-2002 publications.

### 7.3. THE BAND-WAGON EFFECT

Perhaps the concentration towards the HSTKP value is largely influenced by selection effects, for which I will use the colloquial term the 'band-wagon effect'. The band-wagon effect in this case is the effective peer pressure to conform with the widespread popularity of the reasonable HSTKP value. These effects come in many forms: (A) Any study has inputs that may be selected from many different sources, many corrections may be adopted or not adopted, many reasonable cuts on the data could be chosen, and a variety of plausible assumptions may be invoked. It is easy for any investigator to pick and select (consciously or unconsciously) inputs that return values near the popular value. Most of the 31 post-2002 studies report a specific comparison of their new result with prior measures, so the investigators are certainly aware of the relation of their claimed value to the HSTKP 'consensus'. Should an investigator produce a value that is not consistent with the HSTKP value (i.e., |D| is large), then there might be

a strong temptation to select somewhat different (yet still reasonable) input that makes a closer match to the consensus. (B) Should an investigator persist with a result with high |D|, then they are more likely to delay the submission of the paper as they look for flaws or corrections so that their result does not appear simply as a 'poor' measure by comparison with the new community 'standard'. Indeed, with a large value for |D|, some investigators might choose to not submit a paper for publication. (C) Should a paper with a high |D| be submitted, any referee is likely to subject the paper to much closer scrutiny, or perhaps even simply to reject the paper. Thus there exist many and innocuous mechanisms to enforce a band-wagon effect. This interpretation provides an easy explanation for why there was such a sudden and decisive shift around the time of the publication of the HSTKP results.

Unlike with the previous two proposed explanations of the central dilemma, the band-wagon explanation has strong positive evidences: (A) The transition from widely scattered to too-tightly clustered occurs over an unresolved time around 2002 or perhaps 2001-2002. This relatively short duration suggests that the cause of the anomalous clustering is a single event sometime around 2001. The HSTKP publication in 2001 and its widespread publicity correspond in time with the transition. That is, the observed time and duration of the transition is exactly as predicted if the transition is related to the HSTKP. (B) The Cepheid-only median |D| measures follow closely the same behavior over the years as that for all the papers, exactly as predicted if the HSTKP is the cause for the transition to too-tight clustering. (C) The median $|D_{SMC}|$ does *not* get smaller after 2002, exactly as predicted if the LMC |D| transition is associated with 'sociological' effects related to the HSTKP. The reason is that the SMC distance is not a subject of the same peer pressures as the LMC distance because it is not included in the HSTKP consensus.

The existence of strong band-wagon effects is widely known throughout astronomy and physics. The most famous series of recent examples in astrophysics are the groups of researchers who favored particular values for the Hubble constant, where investigators in all 'camps' would repeatedly report sharply concentrated measures of $H_0$. In physics, a famous example is the measure of the electron charge by Robert Millikan, where Millikan's somewhat wrong value was repeatedly confirmed by many subsequent experimenters with only small subsequent shifts, as no one wanted to be at odds with a Nobel prize winner (Feynman 1974). This exact same band-wagon effect is present in the historical measures of many physical constants, where published measures tend to agree with each other until some new value becomes standard, resulting in the best estimate changing over time in a stepwise fashion where each step is significantly different from the preceding step (Yao et al. 2006). The point is that there are ample precedents for the band-wagon effect even amongst the best of scientists.

## 8. CONCLUSIONS

The LMC μ measurements from before the HSTKP are widely scattered with the scatter being much larger than expected based on the reported error bars. After the HSTKP, the 31 LMC μ measurements are tightly concentrated around the HSTKP value of 18.50 mag, with this scatter being much smaller than expected based on the reported error bars. The post-2002 deviation from a Gaussian distribution of errors is significant at the >3-sigma confidence level. The date of this transition is unresolved and around the year 2002 or perhaps 2001-2002.

Three interpretations were considered, although other interpretations might be possible. Correlations between the 31 post-2002 papers are certainly present, but they cannot account for a significant fraction of the sudden post-HSTKP concentration. The over-estimation of error bars is unlikely to provide any substantial amount of concentration. The band-wagon effect has strong positive evidences, including the unresolved duration of the transition, the coincidence in time with the HSTKP, the successful prediction that the Cepheid papers would behave similarly with the whole data set, and the successful prediction that the SMC distances would not suffer the same too-tight clustering. This band-wagon effect has many strong precedents even amongst the best astronomers and physicists from the last decade and century. In all, the evidence strongly points to the cause of the too-tight clustering of the post-2002 LMC distances as being due to a band-wagon effect.

All of these interpretations are worrisome, as they strike at the heart of the measurement process for one of the most important astronomical parameters. So what do I recommend to help this situation? The obvious suggestion is a very close examination of all 31 post-2002 papers for correlations and possible alternative inputs. Ideally, a network containing all plausible sets of assumptions (perhaps with assigned probabilities) can be applied uniformly to the data from the 31 papers so as to produce a single $\mu$ value based on all the data along with error bars that reflect the true range of reasonable inputs. However, such a large analysis is beyond the scope of this paper or indeed of any individual researcher. As a start, some group of knowledgeable experts might take all the calibration data and apply it to all available LMC data for one particular method, say the Cepheids or the RR Lyrae stars. With such comprehensive analyses for the various methods, they can then be combined to produce some final value with a realistic error bar. In the meantime, individual researchers should report their own analysis in such a way that correlations and assumptions are explicitly given. Also, to avoid selection effects, future papers should quantitatively detail the effects on the derived $\mu$ of all available data sources, all plausible data cuts, and all reasonable input assumptions in all combinations. Alternatively, a researcher could *in advance* specify the experimental choices in complete detail (hopefully without regard for how these selections will change the answer) and then stick with these choices all the way to the end.

The purpose of this paper is to alert workers in the field to the problem. Part of this alert is simply to let us know that the current best estimate of $\mu$ is more like the HSTKP value of 18.50±0.10 mag, instead of something like 18.50±0.02 mag. Another part of this alert is to encourage groups of experts to synthesize a global approach for individual methods as well as to combine the various methods. A final part of this alert is to encourage all LMC-distance workers to draft their future papers either with all selections specified rigidly in advance or with the derived $\mu$ values for all reasonable combinations of selections explicitly given in the paper.

I thank Chris Britt and the referee for help with this paper.

# TABLE 1
## Published LMC Distance Measures after the HSTKP

| Date | Paper | Method | $\mu \pm \sigma$ | $D=(\mu-18.50)/\sigma$ |
|---|---|---|---|---|
| 2001.7 | Fitzpatrick et al. 2002 | Ecl. Binary | 18.52 ± 0.05 | 0.40 |
| 2001.8 | Salaris and Groenewegen 2002 | Ecl. Binary | 18.42 ± 0.24 | -0.33 |
| 2002.0 | Bono et al. 2002a | Cepheid | 18.53 ± 0.05 | 0.60 |
| 2002.2 | Mitchell et al. 2002 | SN1987A | 18.50 ± 0.20 | 0.00 |
| 2002.3 | Kerber et al. 2002 | Main Seq. | 18.61 ± 0.07 | 1.57 |
| 2002.4 | Di Benedetto 2002 | Cepheid | 18.59 ± 0.04 | 2.25 |
| 2002.4 | Alves et al. 2002 | Red Clump | 18.493 ± 0.044 | -0.16 |
| 2002.4 | Bono et al. 2002b | Cepheid | 18.51 ± 0.11 | 0.09 |
| 2002.5 | Keller and Wood 2002 | Cepheid | 18.55 ± 0.02 | 2.50 |
| 2002.5 | Whitelock 2003 | Miras | 18.64 ± 0.18 | 0.78 |
| 2002.7 | Benedict et al. 2002 | Cepheid | 18.54 ± 0.15 | 0.27 |
| 2002.9 | Clementini et al. 2003 | RR Lyrae | 18.515 ± 0.085 | 0.18 |
| 2002.9 | Clausen et al. 2003 | Ecl. Binary | 18.63 ± 0.08 | 1.62 |
| 2002.9 | Fitzpatrick et al. 2003 | Ecl. Binary | 18.23 ± 0.09 | -3.00 |
| 2003.1 | Salaris et al. 2003a | Red Clump | 18.53 ± 0.07 | 0.43 |
| 2003.4 | Bono 2003 | RR Lyrae | 18.47 ± 0.07 | -0.43 |
| 2003.5 | Hoyle et al. 2003 | Cepheid | 18.51 ± 0.10 | 0.10 |
| 2003.5 | Gratton et al. 2003 | RR Lyrae | 18.50 ± 0.09 | 0.00 |
| 2003.5 | Fouque et al. 2003 | Cepheid | 18.55 ± 0.06 | 0.83 |
| 2003.5 | Salarais et al. 2003b | Red Clump | 18.47 ± 0.06 | -0.50 |
| 2003.7 | Groenewegen and Salaris 2003 | Main Seq. | 18.58 ± 0.08 | 1.00 |
| 2003.9 | Alcock et al. 2004 | RR Lyrae | 18.43 ± 0.17 | -0.41 |
| 2004.2 | Borissova et al. 2004 | RR Lyrae | 18.48 ± 0.08 | -0.25 |
| 2004.3 | Bellazzini et al. 2004 | Tip of RGB | 18.54 ± 0.15 | 0.27 |
| 2004.3 | Feast 2004 | RR Lyrae | 18.48 ± 0.08 | -0.25 |
| 2004.5 | Dall'Ora et al. 2004a | RR Lyrae | 18.52 ± 0.117 | 0.17 |
| 2004.5 | Dall'Ora et al. 2004b | RR Lyrae | 18.47 ± 0.07 | -0.43 |
| 2004.5 | Maio et al. 2004 | RR Lyrae | 18.51 ± 0.085 | 0.12 |
| 2004.6 | Persson et al. 2004 | Cepheid | 18.50 ± 0.05 | 0.00 |
| 2004.8 | Rastorguev et al. 2005 | RR Lyrae | 18.32 ± 0.08 | -2.25 |
| 2004.8 | Abrahamyan 2004 | Cepheid | 18.569 ± 0.117 | 0.59 |
| 2004.9 | Storm et al. 2004 | Cepheid | 18.48 ± 0.07 | -0.29 |
| 2005.0 | Moskalik et al. 2005 | Cepheid | 18.44 ± 0.10 | -0.68 |
| 2005.1 | Marconi and Clementini 2005 | RR Lyrae | 18.54 ± 0.09 | 0.43 |
| 2005.2 | Gieren et al. 2005 | Cepheid | 18.56 ± 0.11 | 0.55 |
| 2005.9 | Clement et al. 2005 | RR Lyrae | 18.46 ± 0.08 | -0.50 |
| 2006.0 | Keller and Wood 2006 | Cepheid | 18.54 ± 0.018 | 2.22 |
| 2006.5 | Storm et al. 2006 | Cepheid | 18.50 ± 0.05 | 0.00 |
| 2006.6 | Macri et al. 2006 | Cepheid | 18.41 ± 0.16 | -0.55 |
| 2006.6 | Sollima et al. 2006 | RR Lyrae | 18.54 ± 0.15 | 0.27 |
| 2006.8 | Testa et al. 2007 | Cepheid | 18.52 ± 0.14 | 0.14 |
| 2007.0 | Benedict et al. 2007 | Cepheid | 18.40 ± 0.05 | -2.00 |

| | | | | | |
|---|---|---|---|---|---|
| 2007.1 | McNamara et al. 2007 | δ Scuti | 18.48 | ± 0.15 | -0.13 |
| 2007.4 | Grocholski et al. 2007 | Red Clump | 18.40 | ± 0.09 | -1.11 |
| 2007.6 | Ngeow and Kanbur 2007 | Cepheid | 18.49 | ± 0.04 | -0.25 |

**TABLE 2**
**Distributions of published values 1997-Present**

| Date | Number | $<\mu>$ | RMS $\mu$ | Median $|D|$ |
|---|---|---|---|---|
| 1997 | 16 | 18.55 | 0.13 | 1.00 |
| 1998 | 21 | 18.33 | 0.17 | 1.78 |
| 1999 | 15 | 18.49 | 0.16 | 0.82 |
| 2000 | 28 | 18.51 | 0.11 | 0.73 |
| 2001 | 9 | 18.38 | 0.10 | 1.00 |
| 2002 | 21 | 18.48 | 0.10 | 0.71 |
| 2003 | 8 | 18.51 | 0.05 | 0.43 |
| 2004 | 11 | 18.48 | 0.06 | 0.27 |
| 2005 | 3 | 18.52 | 0.05 | 0.50 |
| 2006 | 6 | 18.49 | 0.06 | 0.41 |
| 2007 | 3 | 18.46 | 0.05 | 0.25 |
| 1997-1999 | 52 | 18.44 | 0.18 | 1.24 |
| 2000-2002 | 58 | 18.48 | 0.12 | 0.76 |
| 2003-2007 | 31 | 18.49 | 0.06 | 0.43 |

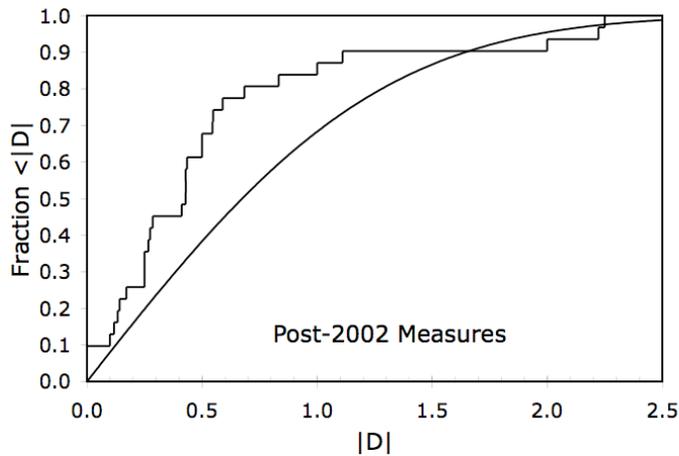

**Figure 1.** Cumulative distributions of |D| from observations and for Gaussian errors. The Kolmogorov-Smirnov (K-S) test is a comparison between the cumulative distributions of |D|, with D=(µ-18.50)/σ, from observations (the stepped curve) and the model (the smooth curve). *If* the published values of the LMC distance modulus (µ) are unbiased and have correctly reported error bars, then the two curves should lie relatively close together. *If* all but a few of the 31 post-2002 values are too-tightly clustered about the HSTKP value of µ=18.50 mag, then the observed curve should step high above the model curve. Indeed, we see that 68% of the published values *are* within 0.5-sigma of the HSTKP value, whereas 68% of the published values *should* be within 1.0-sigma of the HSTKP value. For the K-S test, the maximum deviation between the two curves is 0.33 at |D|=0.59. Such a large deviation is very unlikely (at the 0.0023 probability level, i.e., just over the 3-sigma confidence level) if the published data reports unbiased values with correct error bars. As such, there is a profound problem with the body of the post-2002 measures of the LMC distance. This is the primary result from this paper.

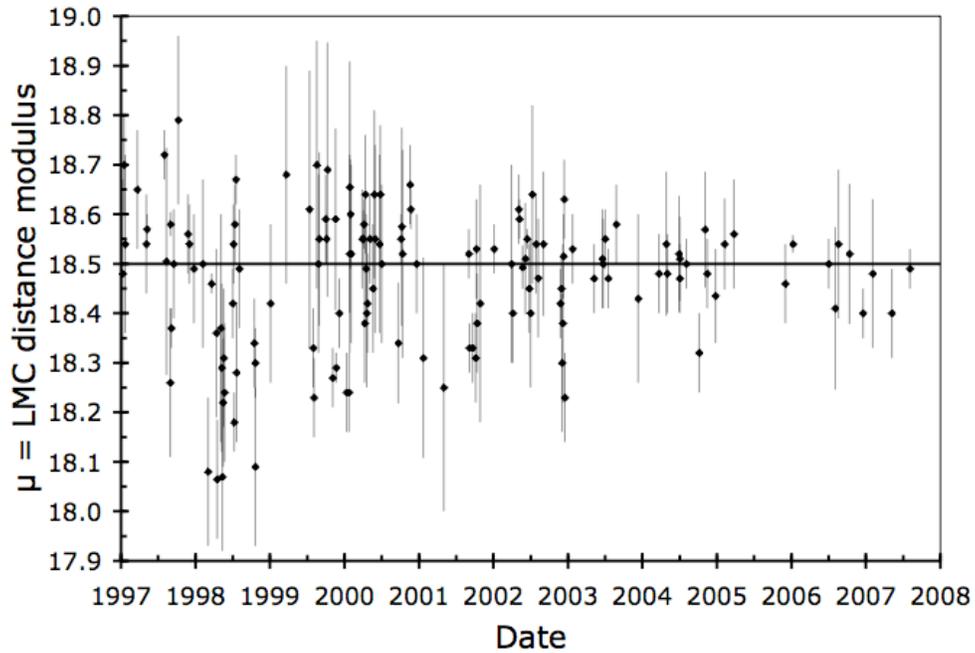

**Figure 2.** Published LMC distance moduli for 1997-present.
The 144 plotted μ values are from Benedict et al. (2002), Clementini et al. (2003), and Table 1 of this paper. The HSTKP value of μ=18.50 mag is represented with a horizontal line. We see that the reported μ values are centered about μ=18.50 mag and they show a distinct narrowing in distribution over the last decade. The reader can easily count from the plot that only 4 out of the 31 post-2002 values have the plotted 1-sigma error bars not intersecting with the HSTKP line. (The same result can be seen in Figure 1, where 87% of the values have |D|≤1.0.) The fact that 87% of the published values are within 1-sigma of the 'consensus' is an alert to our community that something is profoundly wrong with the body of post-2002 LMC distance measures.

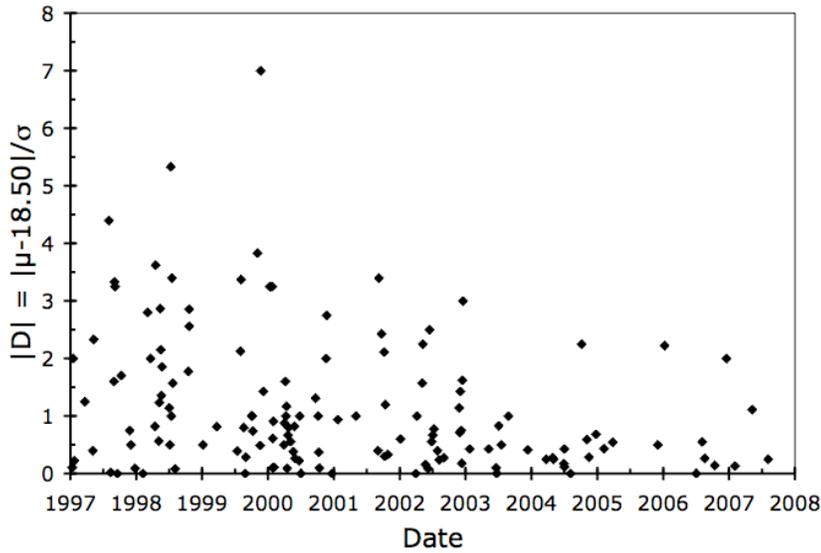

**Figure 3.** |D| for 1997-present.
The vertical axis gives the |D| value for each published LMC distance modulus, with this being the deviation from the HSTKP 'consensus' in units of the quoted error bars. *If the published values are unbiased with the correct uncertainties, then the median |D| should be at 0.68.* For values from before 2000, the median |D| is 1.24, so the scatter is about a factor of 2 times larger than expected for the reported error bars, which is to say that the majority of this pre-HSTKP data had large unrecognized systematic errors. For the values from after 2002, the median |D| is 0.43, which is to say that they are more tightly clustered around the HSTKP than is possible given their quoted error bars. The transition between these two regions appears to be in the year 2002 or perhaps 2001-2002, that is, just after the publication of the HSTKP result with all its massive publicity. As such, it is plausible to associate the cause for the transition to be a result of the HSTKP and the resultant 'consensus' within our community.